\documentclass[traditabstract]{aa}

\usepackage{graphicx}
\usepackage{txfonts}
\usepackage{natbib}
\usepackage{hyperref}
\hypersetup{
  pdftitle={MOLPOP-CEP},
  pdfauthor={Asensio Ramos and Elitzur},
  pdfpagemode=None,        
  colorlinks=true,         
  linkcolor= blue,        
  citecolor= blue,        
  urlcolor=  blue,
  breaklinks=true          
}

\bibpunct{(}{)}{;}{a}{}{,}
\defcitealias{CEP06}{CEP06}

\def\eq#1{\begin{equation} #1 \end{equation}}
\def\eqarray#1{\begin{eqnarray} #1 \end{eqnarray}}
\def\about    {\hbox{$\sim$}}
\def\x        {\hbox{$\times$}}
\def\Dnu      {\hbox{$\Delta\nu$}}
\def\Dv       {\hbox{$\Delta\v$}}
\def\N        {\hbox{$\cal N$}}
\def\half     {\hbox{$1\over2$}}
\def\t(#1){\tau^{#1}_{ul}}
\def\a(#1){\alpha^{#1}_{ul}}

\def\Ie       {\hbox{$I_e$}}
\def\Ji       {\hbox{$\bar J^i_e$}}
\def\E#1{\hbox{10$^{#1}$}}
\def\ga     {\hbox{$\gtrsim$}}
\def\la     {\hbox{$\lesssim$}}

\def\cc     {\hbox{cm$^{-3}$}}
\def\pc     {\hbox{pc}}
\def\kms    {\hbox{km\,s$^{-1}$}}
\def\M      {MOLPOP-CEP}
\def\ncrit  {\hbox{$n_{\rm crit}$}}
\def\CO     {\hbox{CO}}

\font\math = cmmi10
\def\m#1{\hbox{\math \char'#1}} 
\def\v{\m{166}}

\begin{document}

\title{MOLPOP-CEP: An Exact, Fast Code for Multi-Level Systems}
\author{Andr\'es Asensio Ramos\inst{1,2}
   \and Moshe Elitzur\inst{3,4}}

\offprints{aasensio@iac.es}

\institute{ Instituto de Astrof\'\i sica de Canarias,
            38205, La Laguna, Tenerife, Spain: \email{aasensio@iac.es}
            \and
            Departamento de Astrof\'{\i}sica,
            Universidad de La Laguna,
            E-38205 La Laguna, Tenerife, Spain
            \and
            Department of Physics \& Astronomy,
            University of Kentucky, Lexington, KY 40506, USA
            \and
            Astronomy Dept., University of California,
            Berkeley, CA 94720, USA
}
\date{Received ; Accepted}
\titlerunning{MOLPOP-CEP}
\authorrunning{Asensio Ramos \& Elitzur}


\abstract {

We present \M, a universal line transfer code that allows the exact calculation
of multi-level line emission from a slab with variable physical conditions for
any arbitrary atom or molecule for which atomic data exist. The code includes
error control to achieve any desired level of accuracy, providing full
confidence in its results. Publicly available, \M\ employs our recently
developed Coupled Escape Probability (CEP) technique, whose performance exceeds
other exact methods by orders of magnitude. The program also offers the option
of an approximate solution with different variants of the familiar escape
probability method. As an illustration of the \M\ capabilities we present an
exact calculation of the Spectral Line Energy Distribution (SLED) of the CO
molecule and compare it with escape probability results. We find that the
popular large-velocity gradient (LVG) approximation is unreliable at large CO
column densities. Providing a solution of the multi-level line transfer problem
at any prescribed level of accuracy, \M\ is removing any doubts about the
validity of its final results.

}


\maketitle

\section{Introduction}

Much of the information about any astronomical source comes from its spectral
line emission. Lines are the only probe of detailed kinematics, and provide the
tightest constraints on density and temperature. Determining the spectral line
emission from a multi-level system requires solution of the level population
equations for all levels coupled with the radiative transfer equation for every
line connecting them. Because of the complexity and computational demands of
exact solution methods, many analysis codes bypass altogether solution of the
radiative transfer equation, employing instead the escape probability
approximation. This approach requires uniform physical conditions and is
predicated on the conjecture that the effects of radiative transfer can be
lumped into a multiplicative correction to the spontaneous decay rate. This
single ``escape probability" multiplicative factor is supposed to mimic the
effects of radiative transfer in the entire source, and its functional form is
posited from some plausibility arguments. Only the level populations are
considered, calculated from rate equations augmented by these photon escape
factors.

The escape probability approach amounts to an uncontrolled approximation
without internal error estimate because it is not derived from first principles
but instead is founded on a plausibility assumption right from the start. The
only way to assess its error is to repeat the calculation with an exact method
and compare the results\footnote{\cite{dumont03} provide a detailed discussion
of the escape probability method and comparison with ALI calculations.}.
Nevertheless, this inherent shortcoming is often tolerated because of the
simplicity and usefulness of the escape probability approach and the near
impracticality of exact methods. Indeed, current calculations of line emission
from active galactic nuclei, shock fronts and photo-dissociation regions (PDRs)
are only performed in the escape probability approximation, which requires the
underlying physical conditions to be uniform across the emitting region.
Therefore these calculations are forced to replace in each case the variable
conditions with a single value derived from either averaging or from
plausibility arguments that attempt to pick out the most significant region.
Even under these assumptions, the final results are inexact and there is no way
of finding out how large, or how small, the errors involved are. Virtually all
model results reported in the literature are afflicted by these problems.
Furthermore, the escape probability is plainly useless in analysis of spectral
line shapes---this method predicts flat-top profiles for all optically thick
quiescent regions (without large-scale ordered motions), but such sources can
in fact produce double-peaked profiles (\citealt{elitzur_disks12}, and
references therein). Because of this inherent shortcoming, as long as the
analysis is limited to escape probability calculations, much of the available
information cannot be extracted from spectral data. This is an especially
severe handicap when angular resolution is limited and line profiles provide
the only handle on detailed structure.

We have recently developed the Coupled Escape Probability (CEP), a new
radiative transfer method that provides an exact solution for the multi-line
problem while retaining all the advantages of the naive escape probability
approach \citep[][CEP06 hereafter]{CEP06}. In this new technique the source is
divided into zones and formal level population equations, including
interactions with the transferred radiation, are derived rigorously from first
principles. The final formulation does not contain the radiative transfer
equation explicitly, only a set of coupled non-linear level population
equations. These equations are identical in form to those employed in standard
escape probability calculations, but the naive photon escape factors are
replaced by terms derived formally from the exact equations, including those
for the radiative transfer of all lines. These terms introduce coupling between
all zones, and it is this coupling which makes CEP an exact method. Solution of
this set of algebraic equations determines in each zone level populations that
are self-consistent with the line radiation they generate. Once the correct
level populations are derived, line profiles and fluxes are computed from
straightforward summations over zones. The numerical error is controlled by
reducing the zone sizes. Any level of accuracy can be achieved by increasing
the number of zones.

Originally formulated for the slab geometry, the CEP method was subsequently
extended to 3D, first to spherical sources both in hydrostatic equilibrium and
subject to large scale motions \citep{yun09, Yun12}. As in the slab case, the
spherical implementation of CEP yields significant gains in accuracy and
efficiency. \cite{Gersch14} and \cite{Debout16} have further extended CEP to
asymmetrical situations and employed it to model optically thick line emission
from cometary comae. The power of the CEP approach has been recognized also
outside the astronomical community: \cite{ApPhL13} employed CEP to model in
detail extreme-UV laser based lithography, an emerging technology in
high-volume semiconductor chip production. And \cite{ApOpti13} employed CEP in
numerical calculations of the temporal behavior of plasma emission after laser
irradiation.

Since CEP brings a significant speed improvement over other methods, exact
calculations are becoming practical for most line analysis problems. We have
implemented our original method in the code MOLPOP-CEP and it is now
publicly-available (see \S\ref{sec:MOLPOP}). MOLPOP-CEP is a universal line
transfer code that allows the exact calculation of multi-level line emission
from a slab with variable physical conditions for any arbitrary atom or
molecule for which atomic data exist. The code includes error control to
achieve any desired level of accuracy, providing full confidence in its
results.

MOLPOP-CEP has already been employed in numerous studies, including
analysis of the H$_2$D$^+$ resonance line emission from proto-planetary disks
\citep{asensio_h2dp07}; double-peaked line profiles in rotating disks
\citep{elitzur_disks12}; maser observations in supernova remnants
\citep{Pihlstrom14, McEwen14, McEwen16}; far-infrared tracers of oxygen
chemistry in diffuse clouds \citep{Wiesemeyer16}; and ALMA multiple-transition
molecular line observations of an Ultraluminous Infrared Galaxy
\citep{Imanishi17}. The code is available from a dedicated web
site\footnote{\url{https://github.com/aasensio/molpop-cep}} and it can also be
run interactively on a remote
server\footnote{\url{http://www.iac.es/proyecto/inversion/online/molpop\_code/molpop.php}}.

This paper serves to introduce \M\ and its capabilities. For completeness, in
\S2 we describe briefly the CEP radiative transfer formalism and introduce our
notations for the key quantities. Section 3 describes the working of the
program and discusses its implementation. A specific example is provided in
\S4, presenting detailed calculations of CO line emission and comparisons of
the exact results with various variants of the escape probability. The paper
concludes in \S5 with suggested future directions of astrophysical radiative
transfer modeling,

\section{The Line Transfer Problem}

Consider a molecular or atomic multilevel system, with levels $k = 1, 2,
\ldots, L$ ordered by energy. Determining the spectral line emission from this
system requires the coupled solution of the level population equations and the
radiative transfer equations for all the lines connecting them. In the CEP
approach, all radiative quantities are expressed in terms of the level
populations through the formal solution of the radiative transfer equation. As
a result, the complete problem is formulated purely in terms of a set of
non-linear, self-consistent level population equations. Here we summarize
briefly the CEP solution formalism in the slab geometry; for full details, see
\citetalias{CEP06}.

\subsection{CEP Formalism in a Slab}

Consider the plane-parallel geometry, so that physical properties vary only
perpendicular to the surface. The slab is divided into $z$ zones, sufficiently
small that all properties can be considered constant within each zone. The
population per sub-state of level $k$ in zone $i$ ($= 1, 2, \ldots, z$) is
$n_k^i$, where subscripts denote levels and superscripts zones. Then the system
overall population in zone $i$ is
\eq{\label{eq:Norm}
    n^i = \sum_{k = 1}^L g_k n_k^i
}
where $g_k$ is the level degeneracy. Denote by $\ell^i$ the width of the $i$-th
zone and by $\Phi_i(\nu)$ the absorption profile in the zone, normalized
through $\int\Phi_i(\nu) d\nu = 1$ and assumed to have the same functional form
for all transitions. For a transition between lower level $l$ and upper level
$u$ with energy separation $E_{ul} = h\nu_{ul}$ and Doppler width $\Dnu_{ul}^i
= (\nu_{ul}/c)\,\Dv^i$, introduce the dimensionless frequency shift from line
center $x = (\nu - \nu_{ul})/\Dnu_{ul}^i$; the linewidth $\Dv^i$ may reflect
either thermal motions or the dispersion of a micro-turbulent velocity field.
Then the zone optical thickness at frequency $x$ along a ray slanted at $\theta
= \cos^{-1} \mu$ from normal is $\tau_{ul}^{i, i - 1}\Phi_i(x)/\mu$ where
\eq{\label{eq:tau}
    \tau_{ul}^{i, i - 1} = {hc\over4\pi\Dv^i}\,g_uB_{ul}
        \left(n_l^i - n_u^i\right)\ell^i,
}
and $B_{ul}$ is the coefficient of stimulated emission for the transition.
Denote by $A_{ul}$ the corresponding spontaneous transition rate and by
$C_{ul}^i$ the collision rate in zone $i$. Then the CEP level population
equations are
\eqarray{\label{eq:final}
    {dn_k^i\over dt} = -\sum_{l = 1}^{k - 1} A_{kl}p_{kl}^i n_k^i +
      C_{kl}^i\left(n_k^i - n_l^i e^{-E_{kl}/kT^i}\right) \qquad      \\
  \hskip0.2in   +\sum^{L}_{u = k + 1} {g_u\over g_k}
       \left[A_{uk}p_{uk}^i n_u^i +
      C_{uk}^i\left(n_u^i - n_k^i e^{-E_{uk}/kT^i}\right)\right],   \nonumber
}
where $d/dt = 0$ in steady state.  Here
\eq{\label{eq:p_ul}
     p^i_{ul} = \beta^i_{ul} + {1\over\t(i,i-1)}
     \sum_{\stackrel{j = 1}{j \ne i}}^z
     {n_u^j\over n_u^i}\,{n_l^i - n_u^i\over n_l^j - n_u^j}  M_{ul}^{ij}
}
where
\eq{
    M_{ul}^{ij} = -\frac12(\a(i,j) - \a(i-1,j) - \a(i,j-1) + \a(i-1,j-1))
}
and where $\beta_{ul}^i = \beta(\t(i,i-1))$ and $\a(i,j) =
\t(i,j)\beta(\t(i,j))$, with the function $\beta$ defined from
\begin{equation}
\label{eq:beta}
  \beta(\tau) = {1\over\tau}\int_0^\tau \!\!\!dt
                            \int_{-\infty}^{\infty}\!\!\!\Phi(x)dx
                            \int_0^1 \!\!\!d\mu\, e^{-t\Phi(x)/\mu}
\end{equation}
This function was first introduced by \cite{capriotti65}, who also provided
numerical approximations that we found useful for its accurate, efficient
calculation. Plotted in figure \ref{fig:beta_escape} for the Doppler profile,
$\Phi(x) = \pi^{-1/2}\exp(-x^2)$, the function $\beta$ contains the essence of
radiative transfer in the CEP approach. It represents the probability for
photon escape from a slab of optical thickness $\tau$, averaged over the
injected photon direction, frequency and position in the slab. The first term
in the expression defining $p^i$ (eq.\ \ref{eq:p_ul}) is thus the average
probability for photon escape from zone $i$, reproducing one of the common
variants of the escape probability method in which the whole slab is treated as
a single zone (e.g., \citealt{krolik_mckee78}). The subsequent sum in eq.\
\ref{eq:p_ul} describes the effect on the level populations in zone $i$ of
radiation produced in all other zones. Each term in the sum has a simple
interpretation in terms of the probability that photons generated elsewhere in
the slab traverse every other zone and get absorbed in zone $i$, where their
effect on the level populations is similar to that of external radiation.

When external continuum radiation exists, each term in the sums on the
right-hand-side of eq.\,\ref{eq:final} is supplemented by $-B_{ul}\Ji(n_u^i -
n_l^i)$, where \Ji\ is the profile- and angle-averaged intensity of the
external radiation in zone $i$. When the external radiation corresponds to the
emission from dust permeating the source, \Ji\ is simply the angle-averaged
intensity of the local dust emission in the $i$-th zone. When the external
radiation originates from outside the slab and has an isotropic distribution
with intensity \Ie\ ($= J_e$) in contact with the $\tau = 0$ face,
\eq{
  \Ji = \half J_e{1\over\t(i,i-1)}(\a(i,0) - \a(i-1,0)).
}
When the slab is illuminated by parallel rays with intensity \Ie\ ($= 4\pi
J_e$) entering at direction ($\mu_0,\phi_0$) to the $\tau = 0$ face,
\eq{
  \Ji = J_e{\mu_0\over\t(i,i-1)}
        \left[\gamma(\t(i)/\mu_0) - \gamma(\t(i-1)/\mu_0)\right]
}
where
\eq{
   \gamma(\tau) = \int_{-\infty}^{\infty}\left[1 - e^{-\tau\Phi(x)}\right]dx.
}

Eq. (\ref{eq:final}) provides a set of $L - 1$ independent equations for the
$L$ unknown populations in each zone, $n_k^i$. Equation \ref{eq:Norm} for the
overall density in the zone closes the system of equations. It is convenient to
switch to the scaled quantities $n_k^i/n^i$ as the unknown variables and
introduce the overall column density
\eq{
    \N = \sum_{i = 1}^z  n^i\ell^i
}
Neither densities nor physical dimensions need then be specified since only \N\
enters as an independent variable, with the zone partition done in terms of \N\
rather than $\ell$. Apart from external radiation, the problem is fully
specified by three input properties: temperature and density of collision
partners, which together determine the collision terms, and \N, which sets the
scale for all optical depths.

Solution of the set of equations \ref{eq:Norm} and \ref{eq:final} yields the
full solution of the transfer problem for all lines by considering only level
populations; the computed populations are self-consistent with the radiation
field in the slab, including the internally generated diffuse radiation, even
though the radiative transfer equation is not handled at all. Once the
populations are found, radiative quantities can be calculated in a
straightforward manner from summations over the zones. For example, the
contribution of the $u\!\to\! l$ transition to the slab cooling rate per unit
area, accounting for the emission from both faces of the slab, is determined by
its surface flux $F_{\nu,ul}$ and can be characterized by the line cooling
coefficient
\eq{\label{eq:j}
    \textrm{\j}_{ul} \equiv \frac{1}{4\pi\Dnu_{ul}} \int F_{\nu,ul} d\nu,
}
introduced for convenience when \Dnu\ is constant in the slab. Once the level
populations have been determined, \j\ can be calculated from
\eq{\label{eq:jCEP}
\textrm{\j}_{ul} = \half\sum_{i = 1}^z
 \left(\a(i,0) - \a(i-1,0) - \a(z,i) + \a(z,i-1)\right)S^i_{ul}
}
where $S^i_{ul}$ is the line source function in the $i$-th zone.

\subsection{Solution}
\label{sec:Solution}

Solution of the overall system of $z\cdot L$ non-linear algebraic equations
(\ref{eq:Norm}) and (\ref{eq:final}) in all zones determines level populations
that are consistent with the radiation field generated everywhere inside the
slab. In actual numerical calculations, the solution of these equations
provides the exact solution of the radiative transfer problem for all levels
when $\t(i,i-1) \to 0$ for every $i$; the only approximation is the finite size
of the discretization, i.e., the finite number of zones. Therefore to solve the
problem at any desired precision, start with some initial number of zones and
keep refining the divisions until the relative change in level populations
decreases everywhere below the prescribed tolerance.

Because of the non-local nature of radiative transfer, the system of equations
is highly non-linear, making it necessary to apply suitable iterative methods.
We find the multidimensional Newton method most suitable for solution of the
non-linear CEP equations. The efficiency of the Newton method is enhanced in
the CEP approach because the Jacobian matrix can be calculated analytically,
since the functional dependence on population is known explicitly for all
terms. Because the Newton method requires inversion of the Jacobian matrix, the
number of operations in this process increases as the third power of the matrix
dimension and can degrade the performance in cases of very large numbers of
levels and zones. Matrix inversion is avoided in the Bi-CGSTAB iterative scheme
designed by \cite{van_der_Vorst92} for solution of the linear system of the
Newton method. In this scheme, geared toward sparse Jacobian matrices, only the
non-zero matrix elements are stored and used. It is particularly suitable for
the CEP technique because multi-level problems tend to produce sparse matrices,
as each level generally couples to only a limited number of other levels.
MOLPOP-CEP switches automatically to this method when the matrix size exceeds
\E3\x\E3. Matrix inversion can also be avoided by adopting a
$\Lambda$-iteration approach, instead of the Newton method, to solving the set
of non-linear level population equations. As described in \citetalias{CEP06}
(see \S5.2 in that paper), the CEP method is well suited for acceleration
schemes and this approach can be selected in MOLPOP-CEP when the Newton method
slows down.

Whatever the iteration technique, it is always advantageous to start from a
good initial guess. To this end we have implemented two variants of a simple
strategy. The first is to start from the $\N \to 0$ limit with $p = 1$
everywhere, then the level population equations are linear (see
\ref{eq:final}). The solution of this set of linear equations serves as the
initial guess for a low column density \N\ in which all lines are optically
thin. Then \N\ is increased in small steps, with the previous solution taken as
the initial guess for the increased column. The steps are repeated until the
desired column density is reached. The complementary approach is to start from
the $\N \to \infty$ limit, which yields the Boltzmann distribution, use that as
the initial guess for a very large column in which all lines are optically
thick and decrease \N\ toward the target value. Going either way, this
incremental strategy aids convergence and provides the solutions for many
intermediate cases as a byproduct.

\section{The Computer Code MOLPOP-CEP}
\label{sec:MOLPOP}

MOLPOP-CEP is based on a code originally developed by M. Elitzur for solving
the non-LTE level populations of an arbitrary species in the escape probability
approximation. The CEP method has been fully implemented into the original
code, which has been further modularized and ported to Fortran 90. For a slab
with variable physical conditions, MOLPOP-CEP can calculate at a prescribed
level of accuracy the line emission by any atom or molecule for which atomic
data exist. We outline the program capabilities through a brief description of
its input and output.

\subsection{Basic Input}

The input file has a free format, text and empty lines can be entered
arbitrarily. All lines that start with the {\tt `*'} sign are copied to the
output, and can be used to print out notes and comments. This option can also
be useful when the program fails for some mysterious reason, enabling the user
to compare its output with an exact copy of the input line as it was read in
before processing by \M.

A single \M\ run can process an unlimited number of models. To accomplish this,
the program always calls a master input file ({\tt molpop.inp}) containing a
list of the individual cases that are launched sequentially. These individual
models can reside in separate directories, with each model output produced in
the corresponding directory. The information in every input file can be roughly
divided into six groups, which we now describe.

\subsubsection{Radiative Transfer Method}
\label{sec:RT_ver}

\M\ provides a full CEP calculation in the slab geometry. Although the CEP
formalism can handle any profile $\Phi$, currently only the Doppler profile is
implemented. This is sufficient for most cases.

Through the use of keywords, \M\ also offers the choice of a standard escape
probability approximation, included because it provides handling of three
special situations that have not yet been implemented in CEP:
\begin{itemize}
  \item {\em Line overlap}: Under certain circumstances, linewidths become comparable to 
  the separation between different lines so that photons emitted in one transition 
  can be absorbed in another. One commonly affected molecule is 
  OH \citep[e.g.,][]{Guilloteau81}, and this effect has been shown to explain the 
  differences between the emission patterns of OH megamasers and their Galactic 
  counterparts \citep{Lockett08}. The effect is handled with the method 
  of \cite{Lockett89} and can be turned on with a proper flag 
  when \M\ is run in the escape probability mode. Other molecules 
  that have line overlap effects, due to the presence of hyperfine 
  structure, are HCN \citep{cernicharo84, gonzalez-alfonso93} 
  and N$_2$H$^+$ \citep{daniel06}. Another instance of line overlap 
  appears when lines from different species coincide in
  wavelength \citep[e.g.,][]{cernicharo91, cernicharo92, gonzalez-alfonso96}. Such 
  an effect, not currently included in \M, could be implemented with relatively 
  minor modifications.

  \item {\em Dust contribution}: Molecular gas is mixed with dust, therefore 
  line photons can be absorbed by the dust when it is optically thick at 
  the same wavelength. Such absorption and the corresponding emission affect 
  the line transfer (see, e.g., \citealt{elitzur92}, pp.\ 202--203). These 
  modifications are incorporated in the code and can be turned on by 
  setting the appropriate flag.

  \item {\em Maser saturation}:  Transitions with inverted 
  population ($n_u > n_l$) produce maser radiation, which can 
  strongly affect the population inversion when the maser becomes 
  saturated. The saturation effect, strongly dependent on the 
  geometry because of maser beaming, can be described with the 
  escape probability approach \citep{Elitzur90b}. In \M, saturation 
  can be either neglected (the unsaturated maser domain) or handled 
  approximately with the escape probability in a rudimentary manner 
  that ignores maser beaming; a proper treatment of maser saturation requires 
  a more accurate handling of the beaming effect \citep[e.g.,][]{daniel13}.
\end{itemize}

When run in the escape probability mode, \M\ offers four variants of this
approximation. The first is a uniform, static slab, with the escape probability
$\beta$ given in eq.\ \ref{eq:beta}. While this is exactly the same as
performing a CEP calculation with a single zone, the special effects just
described are only available when the escape probability option is selected.
Another variant is the escape probability from a uniform, static sphere
\begin{equation}
\label{eq:sphere}
    \beta_\mathrm{sphere}(\tau)=\frac{1.5}{\tau}
    \left[1 - \frac{2}{\tau^2} +
    \left(\frac{2}{\tau} + \frac{2}{\tau^2}\right)\mathrm{e}^{-\tau}\right]
\end{equation}
where $\tau$ is the optical depth across the diameter (\citealt{van-der-Tak07};
note that this is the escape probability used in their on-line RADEX code). The
two final variants involve approximate solutions for large velocity gradients
(LVG). \M\ implements the escape probability for radial flows in spherical
geometry, with the local velocity variation $\epsilon = d\ln\v/d\ln r$ an input
parameter in the range $0 < \epsilon \le 1$ (see, e.g., \citealt{elitzur92},
pp.\ 39--41). In the literature, the term ``LVG" invariably refers to a
Hubble-type velocity field ($\v \propto r; \epsilon = 1$), which yields
\begin{equation}
\label{eq:LVG}
    \beta_\mathrm{LVG}(\tau)=\frac{1-\mathrm{e}^{-\tau}}{\tau}
\end{equation}
Here the optical depth $\tau$ is given by eq.\ \ref{eq:tau}, with the ratio
$\Dv/\ell$  replaced by the radial velocity gradient $d\v/dr$. \M\ also handles
velocity gradients in plane-parallel geometry, implemented with the
\cite{Scoville74} expression
\begin{equation}
\label{eq:lvg_pp}
    \beta_\mathrm{LVG-PP}(\tau)=\frac{1-\mathrm{e}^{-3\tau}}{3\tau},
\end{equation}
where in this case the relevant velocity gradient is taken normal to the plane.

\begin{figure*}
\includegraphics[width=\textwidth]{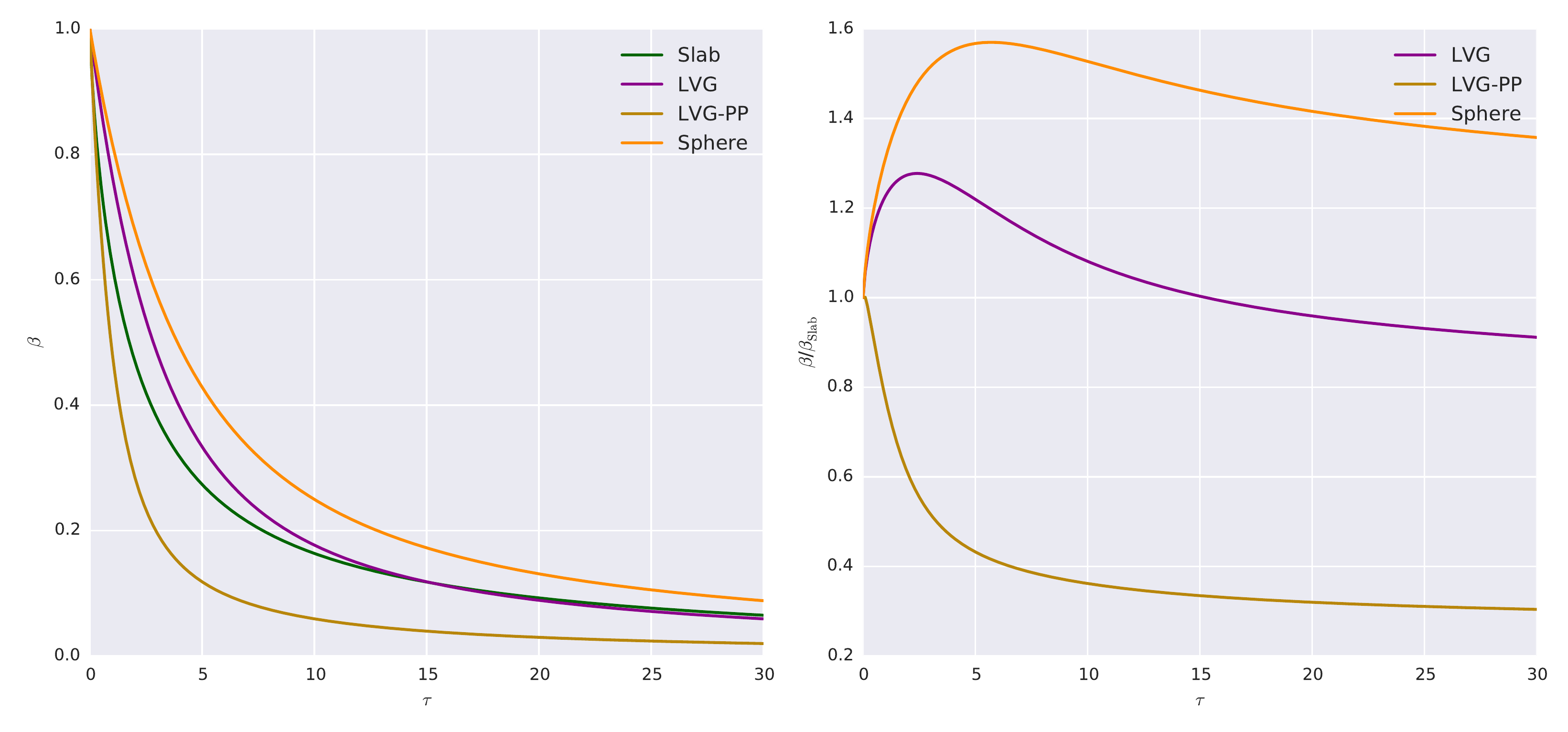}

\caption{{\em Left}: Optical depth variation of $\beta$ (eq.\ \ref{eq:beta}),
plotted in green for the Doppler profile; this is the fundamental CEP function
describing the average probability for escape from a uniform slab with optical
depth $\tau$ across its full width (see text). Also plotted are escape
probability approximations for a uniform, static sphere (eq.\ \ref{eq:sphere})
and LVG approximations for spherical (eq.\ \ref{eq:LVG}) and plane-parallel
(eq.\ \ref{eq:lvg_pp}) geometries. In the LVG cases, the appropriate velocity
gradient replaces the ratio $\Dv/\ell$ in the definition of optical depth. {\em
Right}: The ratio of each escape probability approximation and the fundamental
slab function.} \label{fig:beta_escape}
\end{figure*}

Figure \ref{fig:beta_escape} plots the various escape probabilities as
functions of optical depth and compares them with the slab fundamental
function. The differences from the slab case can be large. Even for the LVG-PP
case, that uses the same geometry, the ratio
$\beta_\mathrm{LVG-PP}/\beta_\mathrm{slab}$ is significantly different from
unity at all $\tau > 0$. Furthermore, the ratio keeps decreasing as $\tau$ is
increasing, as do the ratios for other escape probabilities. The reason is that
at large $\tau$, the various escape probability approximations vary as $1/\tau$
while the proper escape probability for a slab decreases more slowly as
$\sqrt{\ln\tau}/\tau$. The plausibility arguments employed to derive the
various escape probability approximations are too crude to capture the
logarithmic dependence of radiative transfer, rendering them increasingly
unreliable as $\tau$ increases.

\subsubsection{The Molecule/Atom}

An atomic or molecular species is defined by the energy levels, statistical
weights and A-coefficient tabulated in an ordinary text file. Each data file is
identified by its name and directory, allowing for comparison of different sets
of data for the same species. The \M\ distribution package provides a set of
entries from the Leiden Atomic and Molecular Database
(LAMDA)\footnote{\url{http://www.strw.leidenuniv.nl/~moldata}}, and it also
includes a tool that downloads and updates the \M\ input files with the latest
entries from this database. As an example of multiple databases, we include
also sample data files from the Basecol\footnote{\url{http://basecol.obspm.fr}}
database. Installing additional species requires only expansion of the \M\
atomic database; the code itself remains untouched.

\subsubsection{Collisions}

\M\ allows for collisions with up to ten collisional partners. Tabulated
collision rates are invoked through their file names. Since collisional data
are generally available only for a small set of temperatures, \M\ offers a
number of different interpolation methods. Because rate coefficients are not
always available, the code offers some simple analytic approximations (e.g.,
hard sphere collisions) that can be invoked with keywords.

\subsubsection{Physical conditions}

Uniform physical conditions (density, temperature and molecular abundance) can
be specified in the input file. This is the only option when \M\ is run in the
escape probability mode. The full CEP mode allows also variable conditions in
the slab, in which case the spatial profiles of the physical parameters have to
be tabulated in a separate file. The number of CEP zones must then be at least
as large as the number of entries in these profiles.

\subsubsection{External Radiation}

The cosmic blackbody radiation at a temperature of 2.7 K is always included.
Additional radiation fields can also be specified, and in the slab case it is
possible to illuminate each side with different radiation. The external
radiation can include an arbitrary number of diluted blackbody components, each
parameterized in terms of its temperature and dilution factor; this option
covers all cases of illumination by nearby stars. One can also specify an
arbitrary number of radiation fields of the form $(1 -
e^{-\tau_\lambda})B_\lambda(T_d)$, where $B$ is the Planck function and
$\tau_\lambda$ optical depth with the spectral variation of interstellar dust.
This can be used as a crude approximation for emission by dust, characterized
by its temperature $T_d$ and optical depth at visual, that is mixed with the
gas. Additionally, it is possible to specify a radiation field with an
arbitrary spectral shape through a tabulation in a file. Such tabulations can
be generated by some other radiative transfer code such as, e.g., DUSTY
\citep{DUSTY}.

\subsubsection{Numerics}

\M\ offers great control over all aspects of its operation, including various
accuracy parameters. In the multi-zone case, the solution technique can be
chosen between Newton method with analytical derivatives (both matrix inversion
and non-inversion) and an accelerated $\Lambda$-iteration. Convergence is
tested with
\begin{equation}
  R_c(\mathrm{itr}) = \max \left|
          \frac{\mathbf{n}(\mathrm{itr})-\mathbf{n}(\mathrm{itr}-1)}
               {\mathbf{n}(\mathrm{itr})} \right|
\end{equation}
where $\mathbf{n}(\mathrm{itr})$ refers to a vector containing populations for
all the levels in all zones at iteration number itr. Convergence is attained
when $R_c({\rm itr})$ decreases below an input-specified threshold. For safety,
a maximum number of iterations is also specified to stop execution in case of a
runaway calculation. An orderly completion of a run occurs when the column
density reaches the maximum/minimum column specified in the input for the
scheme that iteratively increases/decreases the column density to improve the
convergence properties (\S\ref{sec:Solution}). If the CEP method is used, it is
also possible to turn on the grid convergence. The problem is solved in grids
of increasing number of zones until the maximum relative change between two
consecutive grids decreases below the user-defined threshold.


\begin{figure*}
\includegraphics[width=.95\textwidth]{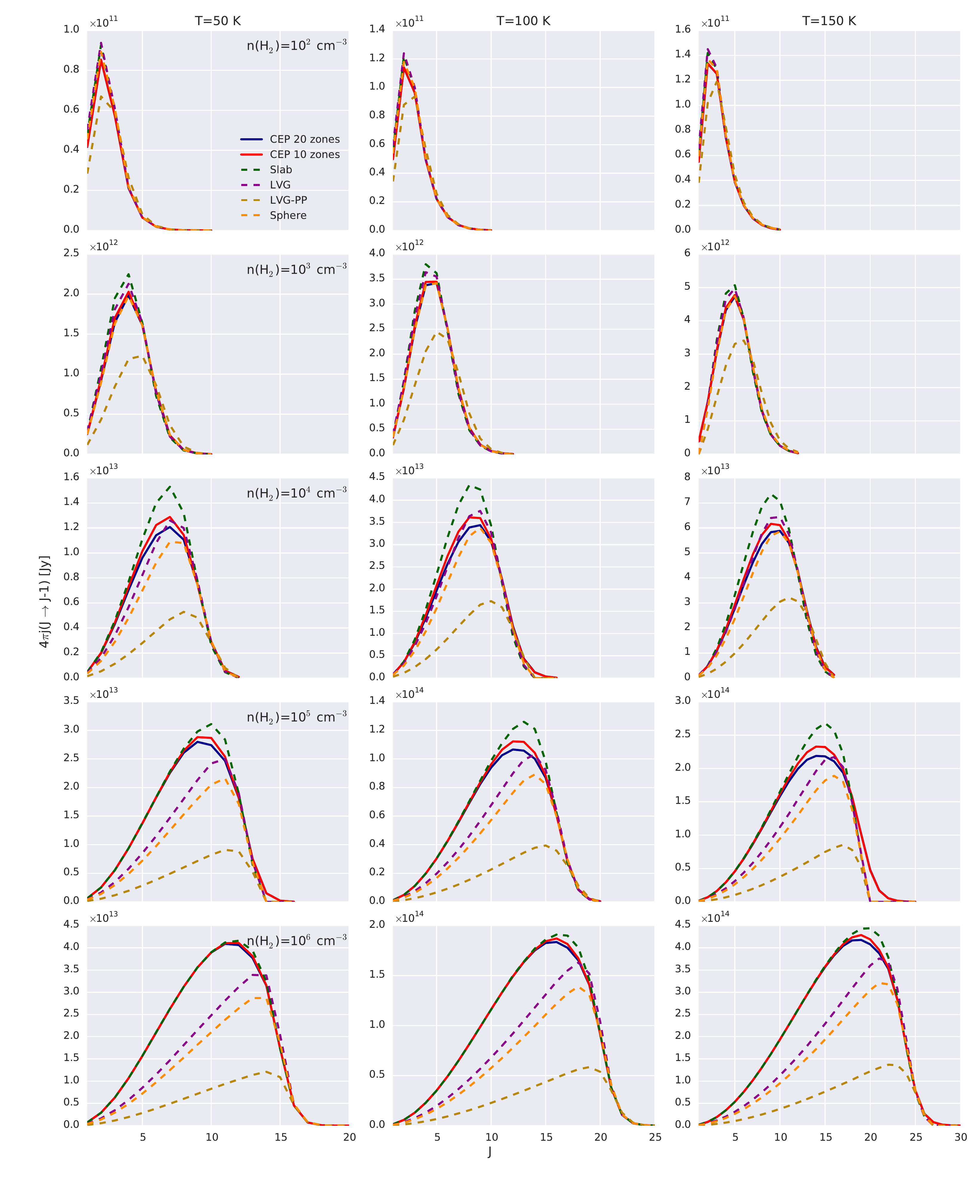}

\caption{\M\ results for the CO SLEDs of uniform slabs, computed with different
methods (the line cooling coefficient $\textrm{\j}$ is defined in eq.\ \ref{eq:j}). The
$y$-axis scale of each panel is listed at its top-left corner. Columns have the
same temperature, listed at the top, rows the same H$_2$ density, marked in
each leftmost panel. Optical depths are set from $[\CO]\,\ell/\Dv$ = \E{-4}
pc\,(\kms)$^{-1}$, where [CO] is the CO abundance. Full CEP calculations using
grids with 10 and 20 zones are plotted with solid lines of different colors, as
marked. When the 20-zones line (blue) is missing, it is because it coincides
with the 10-zones result (red), demonstrating CEP convergence in practically
all cases. Dashed lines show results for different escape probabilities (see
figure \ref{fig:beta_escape}). Note that the slab escape probability
calculation is the same as single-zone CEP.
} \label{fig:SLED-4}
\end{figure*}

\subsection{Output and its Control}

A single \M\ run generates prodigious amounts of data, more than might be
needed in any particular application of the code. Various on/off switches
provide control over the output. The numerous options are described in detail
both in the manual and in the sample input files provided with the \M\ package.
As noted above, every input line that starts with {\tt *} is echoed in the
output, making is easy to incorporate notes and comments.

\section{An Illustrative Example}

As a demonstration of the capabilities of \M\ we present here a detailed
analysis of CO emission, calculating what has become known as its Spectral Line
Energy Distribution \citep[SLED;][]{Weiss05, Papadopoulos12}. Several
transitions of the CO rotational ladder are observable in high-redshift
galaxies \citep[e.g.,][]{solomon_molecular_highz05}. Analysis of diagrams that
plot line emission vs. rotational quantum number, i.e., CO SLED, offers one of
the most direct diagnostic techniques to study the excitation conditions of the
molecular gas. In LTE, the level populations depend only on the local
temperature $T$, while the line emission depends additionally on the CO column
density per unit bandwidth. In that case, the shape of the CO SLED normalized
to the $J = 1 \to 0$ emission uniquely determines $T$, and the CO column
follows directly from the magnitude of the measured flux. Outside of LTE, the
level population distribution is determined through the competition between
collisional and radiative interactions, bringing in also the local H$_2$ number
density and necessitating a full radiative transfer calculation.

To our knowledge, the CO SLED has only been computed in the LVG escape
probability approximation (eq.\ \ref{eq:LVG}). Inherently, such calculations
cannot be exact even when the physical conditions are uniform: The strength of
radiative interactions varies with distance to the surface for optically thick
lines, therefore the population distribution of transition levels does depend
on position in the source. This effect cannot be captured by the escape
probability approach, where the entire source is described by a single
excitation temperature. Furthermore, every line on the CO rotational ladder has
a different optical depth, therefore they are affected differently by this
variation. As a result, the reliability of studies based on SLED analysis
\citep[e.g.,][]{Papadopoulos12, daCunha13} cannot be assessed because the
escape probability method does not contain an error estimate so it is
impossible to know whether its deviations from the correct results fall within
the observational uncertainties. An example where the conclusions based on
escape probability analysis are in serious error is the ratio of the two $^3$P
oxygen cooling lines at 63 and 145 $\mu$m \citepalias[see][]{CEP06}. As is
always the case, it is impossible to know whether the escape probability
results are meaningful without comparison with an exact calculation.

Here we perform such a comparison for the CO SLED. For a meaningful comparison,
one must utilize the same computed quantity. While the CEP method provides
rigorous expressions for all radiative quantities, the escape probability
approximation can be expected to produce useful results only in some average
sense. This makes the line cooling coefficient \j\ (eq.\ \ref{eq:j}) especially
useful. In the case of a slab, the CEP formalism yields eq.\ \ref{eq:jCEP}.
Since a single-zone CEP calculation is equivalent to a slab escape probability
calculation, we can read off immediately the proper escape probability
expression by inserting $z = 1$. This yields the familiar escape-probability
result $\textrm{\j}_{ul} = h\nu_{ul}A_{ul}\beta_{ul}N_{u}/4\pi\Dnu_{ul}$, where $N_u$ is
the column density of the upper level (e.g., \citealt{elitzur92}, p 28).

We have conducted an extensive set of \M\ calculations for a uniform slab,
varying the temperatures from 50 K to 150 K and the molecular hydrogen density
from \E2\ to \E6\,\cc. The slab thickness, linewidth and CO abundance, $[\CO] =
n(\CO)/n(\rm H_2)$, enter only through the combination $\gamma =
[\CO]\,\ell/\Dv$; the scale of all optical depths is controlled by the product
$\gamma n$ (eq.\ \ref{eq:tau}). Beginning with \cite{Weiss05}, CO SLED
calculations assumed a constant value for this parameter, and for compatibility
we adopt this assumption, presenting results for $\gamma = \E{-4}$ and \E{-5}
pc\,(\kms)$^{-1}$; these values bracket the likely range in local and high-$z$
galaxies \citep{weiss_co_sed07}. Collision rates are from \cite{Yang10} with
ortho-to-para H$_2$ ratio of 3. The only external radiation is the cosmic
microwave background.

\subsection{Exact Solutions}
\label{sec:exact}

\begin{figure*}
\includegraphics[width=0.95\textwidth]{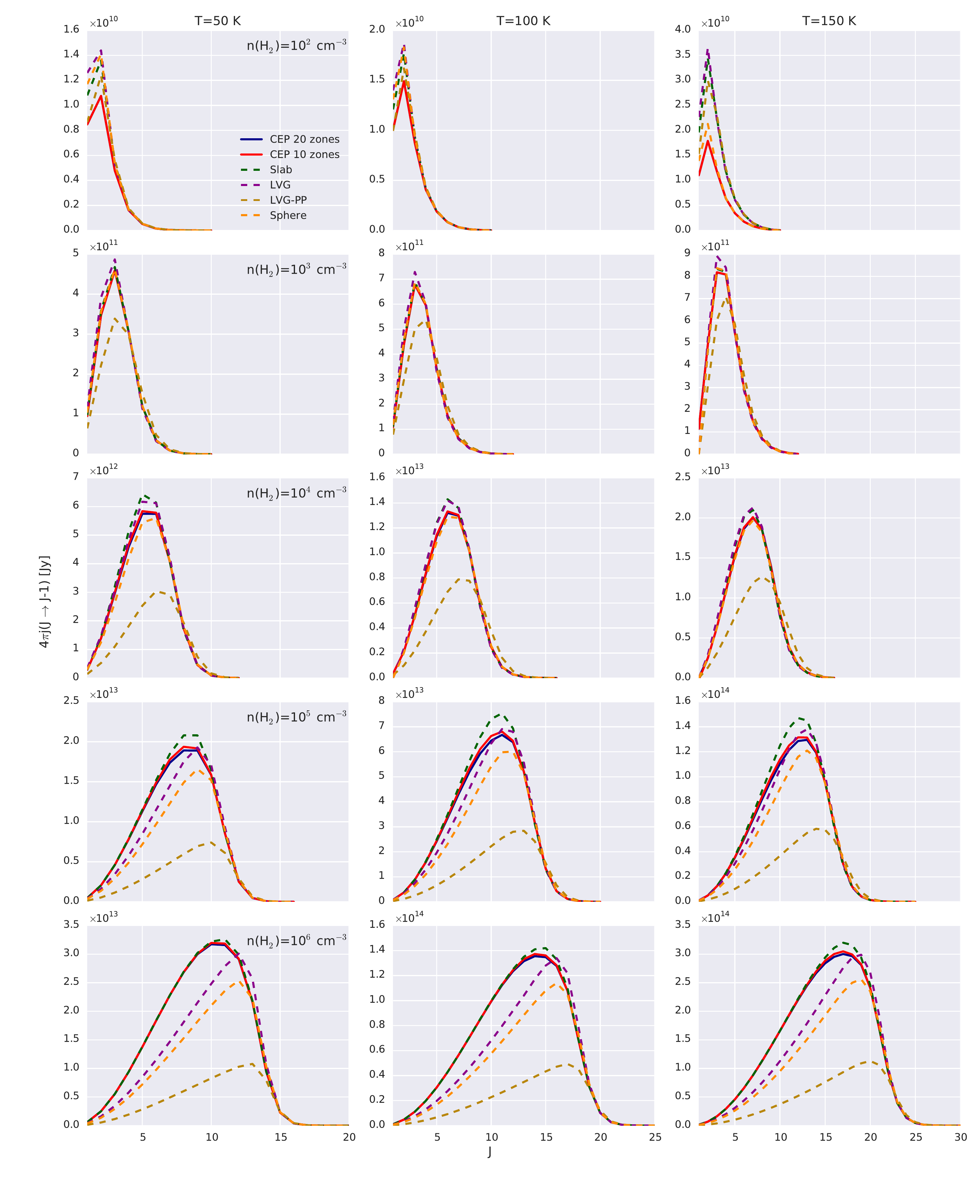}

\caption{Same as figure \ref{fig:SLED-4}, only $[\CO]\,\ell/\Dv$ = \E{-5}
pc\,(\kms)$^{-1}$.
} \label{fig:SLED-5}
\end{figure*}

Figures \ref{fig:SLED-4} and \ref{fig:SLED-5} show the resulting SLEDs in terms
of $4\pi\textrm{\j}$, the line cooling coefficient in units of flux density, for the two
studied values of $\gamma$. Density varies among panels in each column,
temperature between the columns. The exact solutions are shown with solid blues
lines; we have verified that in all cases the CEP calculations with 20 zones
are fully converged---increasing further the number of zones does not change
the results. It is interesting to note how different the actual SLEDs are from
LTE predictions. If all models were in LTE, the SLEDs would be identical in
each column and would differ only between the columns. The actual behavior is
just the opposite: CO SLEDs vary only moderately with temperature in the
displayed range but are strongly dependent on density. Also, some combinations
of $T$ and $n(\rm H_2)$ produce inversions in low $J \to J - 1$ transitions, as
noted already by \cite{Goldreich74}. These cases were ignored in the figures.


Each panel also shows CEP calculations with 10 zones (solid red lines).
Doubling the number of zones from 10 to 20 hardly affects the results, in many
cases the corresponding plots are barely distinguishable from each other. In
figure \ref{fig:SLED-4}, where $\gamma = \E{-4}$, the largest differences from
the exact solution are at $n(\rm H_2)$ = \E4\,\cc, declining towards both
smaller and larger densities. Even though CO is a multi-level system, these
trends can be understood from the solution of the two-levels problem: As shown
in \citetalias{CEP06}, the largest deviations of the slab escape probability
(i.e., single zone) calculation from the exact solution of the two-levels
problem occur around $\tau n/\ncrit \sim 1$, where \ncrit\ is the transition
critical density. When $\gamma = \E{-4}$, this condition is fulfilled around
the SLED peak at about $n(\rm H_2)$ = \E4\,\cc, and that is where the largest
deviations occur from the exact solutions. Since $\tau \sim \gamma n$ (eq.\
\ref{eq:tau}) and $\gamma$ is held fixed, $\tau n \propto n^2$ along each
column, and the deviations from the exact solution are rapidly diminished
moving away from $n(\rm H_2)$ = \E4\,\cc\ in either direction. In Figure
\ref{fig:SLED-5}, where $\gamma = \E{-5}$, optical depths are smaller and the
10-zone results are practically identical to the exact ones in all cases. Thus
the CEP calculations with only 10 zones are practically convergent in all the
presented solutions.

\subsection{Escape Probability Calculations}

Solutions obtained with various escape probability approximations are shown
with dashed lines. In figure \ref{fig:SLED-4}, virtually all escape probability
approximations reproduce rather well the exact solutions in the top two rows,
where optical depths are the smallest. The sole outlier is the slab LVG
approximation (LVG-PP; eq.\ \ref{eq:lvg_pp}). Even though this approximation
purports to describe the very same slab geometry, it fails to reproduce its
proper solution even at the smallest density, and its deviations from the exact
results keep increasing continuously with density (and optical depths).
Evidently, the plausibility arguments that yielded this approximation
\citep{Scoville74} were off; the plane-parallel LVG escape probability is
inadequate for CO radiative transfer calculations .

The slab escape probability approximation (eq.\ \ref{eq:beta}), equivalent to a
single-zone CEP calculation, is doing much better, producing reasonably good
agreement with the exact solution. Similar to the 10-zone CEP calculation, and
for the same reasons (\S\ref{sec:exact}),  the largest deviations from the
exact solution occur around $n(\rm H_2)$ = \E4\,\cc\ when $\gamma = \E{-4}$
(fig.\ \ref{fig:SLED-4}). Yet even in this case the \hbox{1-zone} calculation
is within \about 27\% of the exact solution, and its deviations decrease
towards both lower and higher densities. When $\gamma = \E{-5}$ (fig.\
\ref{fig:SLED-5}), the slab escape probability similarly provides an adequate
approximation in all cases except for the lowest density of \E2\,\cc, where the
discrepancy is more than a factor of 2. Since $\gamma$ is a factor of 10 lower
than in fig.\ \ref{fig:SLED-4}, these models with $n(\rm H_2)$ = \E2\,\cc\
stand out as having the lowest optical depths of all the cases presented here,
implying that large discrepancies occur only at low optical depths. All in all,
the results show that the slab escape probability provides an excellent
approximation for CO SLEDs in all uniform slabs with $n(\CO)\ell/\Dv\ \ga\
\E{-3}\,\cc\,\pc\,(\kms)^{-1}$.

The figures also show the results of calculations with the two escape
probabilities for spherical geometry offered by \M: a uniform, static sphere
(eq.\ \ref{eq:sphere}) and LVG spherical expansion (eq.\ \ref{eq:LVG}). Both
versions provide excellent approximations for the exact results at densities
$n(\rm H_2) \le $\E4\,\cc\ when $\gamma = \E{-4}$. Beyond that,  the deviations
from the exact solution increase with density (and optical depths), with the
SLED peaks occurring at the wrong $J$. These large deviations cannot be
attributed to the effect of a different geometry on radiative transfer---the
two escape probabilities are successful at $n(\rm H_2) = $\E4\,\cc, where
optical depths are already significant. Rather, these versions of $\beta$ are
afflicted by the same weakness as the LVG-PP escape probability, which produces
the same SLED peaks: all three yield $\beta \sim 1/\tau$ at large $\tau$,
failing to reproduce the additional logarithmic variation, which becomes
important when $\tau$ is sufficiently larger. The various escape probability
approximations are too crude to capture the logarithmic dependence of radiative
transfer, losing their reliability at large optical depths. As seen in Fig.
\ref{fig:SLED-5}, all escape probabilities are off at $n(\rm H_2)$ = \E2\,\cc\
when $\gamma = \E{-5}$. Therefore, the common LVG escape probability, as well
as the static-sphere escape probability, provides an adequate approximation for
CO radiative transfer only in the range \E{-3}\,\cc\,pc\,(\kms)$^{-1} \la\
n(\CO)\ell/\Dv$ \la\ \E{-1}\,\cc\,pc\,(\kms)$^{-1}$; the utility range of the
two spherical escape probabilities is limited both at small and large CO column
densities.

\section{Discussion}

High-resolution facilities such as ALMA enable detailed observations that 
can only be understood through realistic 3D modeling. However, on top of complexity 
and heavy computational demands, proper definitions of 3D models involve a large 
number of uncertainties. While such models are clearly necessary in many cases, 
the widespread success of the RADEX code (as of this writing, \citealt{van-der-Tak07} 
has received more than 700 citations) shows there remains a large demand for 
efficient, if simple, radiative transfer tools. \M\ offers such a tool, providing 
an efficient, exact and verifiable solution of radiative transfer.

While species such as N$_2$H$^+$, HCN, HCO$^+$, CS or NH$_3$ are required 
for probing densities higher than about \E4\,\cc, the two most common 
indicators of physical conditions in molecular clouds and PDRs are CO, whose 
SLED analysis is presented here, and O{\small I} cooling lines, 
studied earlier in \citetalias{CEP06}. In both cases LVG escape probability 
calculations can produce erroneous results, therefore this widespread technique 
is an unreliable analysis tool. Although the escape probability is occasionally 
adequate in some limited regions of parameter space, it is impossible to discern 
any trends or establish guidelines that would enable reliance on it alone for 
calculating line intensities for multi-level systems. The only way to get 
the correct result is to perform a correct calculation; there is no viable 
shortcut through the escape probability approximation. \M\ eliminates the 
guesswork, offering a public tool for exact solution of the line transfer 
problem. While the current version handles only slabs, this geometry 
suffices for accurate modelling of both shock- and photodissociation 
fronts, covering the needs of many interstellar studies.

\begin{acknowledgements}
We thank Paul Goldsmith, Massa Imanishi and Jos\'e Cernicharo, the referee. 
for useful comments and suggestions. AAR acknowledges financial support by the Spanish 
Ministry of Economy and Competitiveness (MINECO) through project AYA2014-60476-P.
\end{acknowledgements}


\end{document}